\documentclass[12pt]{iopart}
\usepackage{amssymb,epsfig,graphicx,adjustbox}

\begin{document}

\title[Strain glass versus antisite disorder]{Strain glass versus antisite disorder induced ferromagnetic state in Fe doped Ni-Mn-In Heusler martensites}

\author{R. Nevgi$^1$, K. R. Priolkar$^1$ and M. Acet$^2$ }
\address{$^1$Department of Physics, Goa University, Taleigao Plateau, Goa 403206 India}
\address{$^2$Experimentalphysik, University of Duisburg-Essen, 47048 Duisburg, Germany}
\ead{krp@unigoa.ac.in}
\vspace{10pt}
\begin{indented}
\item[]October 2020
\end{indented}

\begin{abstract}
Fe doping in Ni$_2$Mn$_{1.5}$In$_{0.5}$ results in suppression of the martensitic phase via two contrasting routes. In Ni$_2$Mn$_{1.5-x}$Fe$_{x}$In$_{0.5}$, the martensitic phase is converted to a strain glassy phase, while in Ni$_{2-y}$Fe$_y$Mn$_{1.5}$In$_{0.5}$, a cubic ferromagnetic phase results at the expense of the martensite. Careful studies of magnetic and structural properties reveal the presence of the impurity $\gamma -$(Fe,Ni) phase as the reason for the emergence of non-ergodic strain glassy phase when Fe is sought to be doped at Y/Z (Mn) sites of X$_2$YZ Heusler alloy. Whereas attempts to dope Fe in the X (Ni) sublattice result in an A2 type antisite disorder that promotes a ferromagnetic ground state.
\end{abstract}

\maketitle

\section{Introduction}

The martensitic transition in Heusler alloys is predominantly controlled by the addition of impurities resulting in numerous applications. Some of the fascinating properties which originate as a result of doping include large magnetocaloric effect \cite{Krenke42005,Moya752007}, giant barocaloric effect \cite{Manosa92010}, large magnetoresistance \cite{Koyama892006,Sharma892006,Yu892006} and magnetic field induced strains \cite{Kainuma4392006}. The transformation behaviour largely depends on the type and concentration of impurities. In Ni-Mn-Z (Z = In, Sn, Sb) Heusler alloys, structural transformation from cubic L$2_1$ austenite to lower symmetry martensite or vice-versa occurs when the dopant concentration exceeds a critical value. This is associated with the alteration of magnetic ground state from ferromagnetic to non-magnetic/antiferromagntic \cite{Krenke732006,Priolkar872013} leading to some interesting phenomena like kinetic arrest \cite{Ito922008,Sharma762007}, exchange bias \cite{Khan912007}, as well as emergence of non-ergodic states like magnetic cluster glass \cite{Yadav92019,Nevgi322020}.

Recent studies have also shown the occurrence of strain glass in Heusler alloys with impurity doping \cite{Wang982012,Nevgi1122018}. Strain glass phase reportedly occurs in all ferroelastic/martensitic alloys beyond a critical dopant concentration \cite{Ren2012}. The presence of a defect phase like body centred Ni phase in Ni rich NiTi alloys, impede the long range ordering of the elastic strain vector leading to a frozen ferroelastic phase. On the other hand, no pre-transition phases are reported in Mn-rich Ni-Mn-In Heusler alloys. The ability of the Heusler structure to accommodate the stress caused by atomic size mismatch between Mn and In atoms is believed to be responsible for the absence of non-ergodic phases \cite{Nevgi322020}. Doping of transition metal like Fe in the magnetic shape memory alloys is known to suppress martensitic transition. In some cases, Fe addition results in an emergence of structural impurity phases \cite{Chen1012012,Tan72017,Zhang2016481,Lobo2014116}. Site selectivity of the dopant atom appears to play a key role in the ground state of the resultant alloy. It is observed that doping Fe at the expense of Mn to realize Ni$_{2}$Mn$_{1.5-x}$Fe$_{x}$In$_{0.5}$ leads to suppression of the martensitic state. At a critical concentration of $x = 0.1$, despite the average structure being martensitic, strain glassy phase appears below the glass transition temperature $T_g$ = 350 K \cite{Nevgi1122018}. On the other hand when Fe is doped to replace Ni, to realize Ni$_{2-y}$Fe$_y$Mn$_{1.5}$In$_{0.5}$, the ferromagnetic interactions are enhanced with a complete suppression of the martensitic state at $y = 0.2$ \cite{Nevgi7972019}.

Therefore, it is essential to understand the role of the dopant in the occurrence of strain glassy phase, in particular the nature of the structural defect created, that leads to the formation of the strain glass. It is also equally important to understand the apparent site selectivity of the dopant atom that results in a cubic ferromagnetic ground state. To that effect, we have studied, in detail, the structural and magnetic properties of Ni$_{2}$Mn$_{1.5-x}$Fe$_{x}$In$_{0.5}$ ($0 \le x \le 0.2$) and Ni$_{2-y}$Fe$_y$Mn$_{1.5}$In$_{0.5}$ ($0 \le y \le 0.2$). The results suggest that in Ni$_{2}$Mn$_{1.5-x}$Fe$_{x}$In$_{0.5}$ where the X sublattice of X$_2$YZ Heusler is fully occupied, dopant Fe segregates into an impurity phase leading to a non-ergodic ground state. Whereas, when Fe is sought to be doped into the X sublattice to achieve Ni$_{2-y}$Fe$_y$Mn$_{1.5}$In$_{0.5}$, it promotes an A2 type antisite disorder by itself occupying Y/Z sublattice and facilitates a cubic ferromagnetic ground state.

\section{Experimental}

The alloys were prepared by arc melting stoichiometric amounts of high purity elements ($>$ 99.9\%)in an argon atmosphere. The homogeneity of the ingots was ensured by flipping the individual ingot a few times during the preparation process. The ingots were cut, powdered, covered in tantalum foil, and were vacuum sealed in quartz tubes to be annealed at 750 $^\circ$C for 48 hours and eventually quenched in ice cold water. The alloy compositions were verified using scanning electron microscopy with energy dispersive x-ray (SEM-EDX) analysis and were found to be within $\pm$2\% of the stoichiometric values. Room temperature x-ray diffraction measurements were performed on the powdered alloys using Mo $K_\alpha$ radiation in the angular range of 10$^\circ$ to 50$^\circ$. Synchrotron x-ray diffraction measurements were carried out on BL-18B at Photon Factory, KEK, Tsukuba, Japan using incident photons of 16 KeV at 300 K and 500 K to trace the structural changes with temperature. Temperature dependent magnetization measurements M(T) were carried out in the temperature range of 10 K to 380 K using a SQUID magnetometer in the applied magnetic field of 5 mT and 5 T  wherein the samples were first cooled in zero applied magnetic field from room temperature to the lowest temperature and the data was recorded while warming (ZFC), the subsequent cooling (FCC) and warming (FCW) cycles. The resistivity measurements were performed using standard four probe method in the temperature interval of 200 K -- 400 K and frequency dependence of AC storage modulus and loss were carried out using Dynamical Mechanical Analyzer (Q800, TA Instruments) as described in \cite{Nevgi1122018}. Extended X-ray Absorption Fine structure (EXAFS) were employed to perform local structural studies at Ni K (8333 eV), Mn K (6539 eV) and Fe K (7111 eV) edges at the P65 beamline (PETRA III Synchrotron Source, DESY, Hamburg, Germany). Using gas ionization chambers as detectors, the incident (I0) and the transmitted (I) photon energies were simultaneously recorded. The thickness $t$ of the absorbers were adjusted so as to obtain the absorption edge jump, $\Delta \mu t \leq 1$. For the Ni and Mn K edge EXAFS data was averaged over three scans collected in transmission mode while and at the Fe K edge ten scans were collected in fluorescence mode to average the statistical noise. The EXAFS data was analyzed using well established procedures in Demeter suite \cite{Raval200512}.

\section{Results}

\begin{figure}[h]
\begin{center}
\includegraphics[width=\columnwidth]{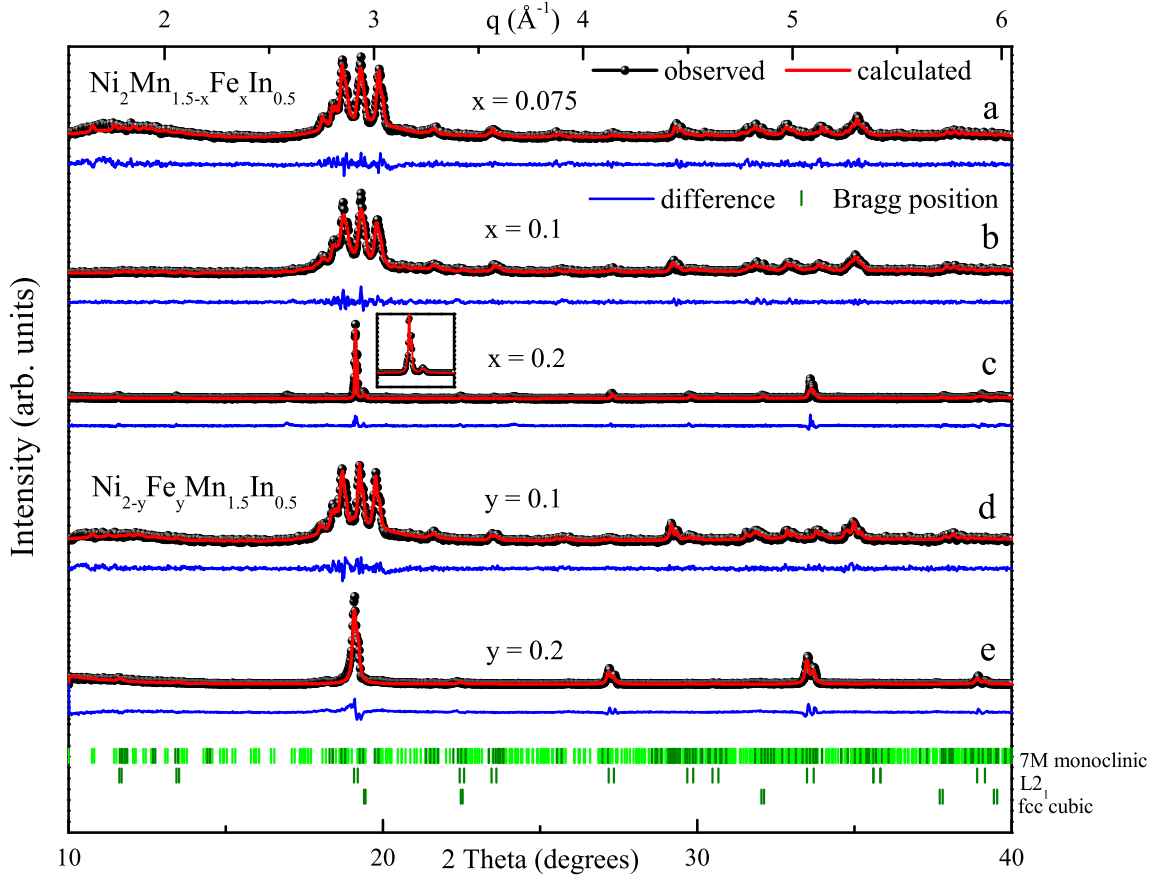}
\caption{The room temperature Lebail refined x-ray diffraction data as the function of two theta and wave vector q for the alloys 0.075 $\leq x \leq$ 0.2 in the series Ni$_{2}$Mn$_{1.5-x}$Fe$_{x}$In$_{0.5}$ (a to c) and for the alloys $y$ = 0.1, 0.2 in the series Ni$_{2-y}$Fe$_{y}$Mn$_{1.5}$In$_{0.5}$ (d and e). Inset highlights the presence of minor fcc cubic phase seen the alloy $y$ = 0.2 (c).}
\label{fig:XRD}
\end{center}
\end{figure}

Fig. \ref{fig:XRD} represents the room temperature x-ray diffraction data analyzed by Le Bail method using Jana 2006 software\cite{Petricek2292014}. In the series Ni$_{2}$Mn$_{1.5-x}$Fe$_{x}$In$_{0.5}$, the alloys with lower Fe concentration ($x \leq 0.1$) exhibit 7M modulated martensitic structure (Fig. \ref{fig:XRD}a and b) just like the parent Ni$_{2}$Mn$_{1.5}$In$_{0.5}$ \cite{Nevgi7972019}. With the increase in Fe content ($x = 0.2$), there is a structural transition from the martensitic phase to a major $L2_1$ and a minor impurity phase, as seen in Fig. \ref{fig:XRD}c. The additional Bragg peaks of the impurity phase can be fitted to a face centred cubic (fcc) phase identified later as $\gamma -$(Fe,Ni) phase. In the second series Ni$_{2-y}$Fe$_{y}$Mn$_{1.5}$In$_{0.5}$, the 7M modulated phase is converted to a cubic Heusler phase with the increase in Fe concentration from $y = 0.1$ to $y = 0.2$ (Fig. \ref{fig:XRD}d and e). The refinement results are summarized in Table \ref{table0}.

\begin{table}[h]
\caption{\label{table0} The refined crystallographic data for the alloy compositions $x$ = 0.075, 0.1 and 0.2 in the series Ni$_{2}$Mn$_{1.5-x}$Fe$_{x}$In$_{0.5}$, and $y$ = 0.1, 0.2 in the series Ni$_{2-y}$Fe$_{y}$Mn$_{1.5}$In$_{0.5}$.}
\setlength{\tabcolsep}{0.3pc}
\vspace{0.3cm}
\centering
\begin{tabular}{|l|l|l |}
\hline
\textbf{Chemical formula} & \textbf{Space group} & \textbf{Lattice parameters}\\[0.5ex]
\hline
Ni$_{2}$Mn$_{1.425}$Fe$_{0.075}$In$_{0.5}$   & $I2/m(\alpha0\gamma)00$ & a = 4.450(5) \AA, b = 5.624(7) \AA, c = 4.376(4) \AA \\
                                             &                         & $\beta$ = 93.93(1)$^\circ$, $q$ = 0.342(5)$c^*$ \\
\hline$\deg$
Ni$_{2}$Mn$_{1.4}$Fe$_{0.1}$In$_{0.5}$       & $I2/m(\alpha0\gamma)00$ & a = 4.438(2) \AA, b = 5.644(1) \AA, c = 4.355(1) \AA\\
                                             &                         & $\beta$ = 93.32(2)$^\circ$, $q$ = 0.337(1)$c^*$\\
\hline
Ni$_{2}$Mn$_{1.3}$Fe$_{0.2}$In$_{0.5}$       & $Fm-3m$ ($L2_1$)          & a = 5.983 (3) \AA\\
                                             & $Fm-3m$ ($\gamma$)        & a = 3.616 (2) \AA\\
\hline
Ni$_{1.9}$Fe$_{0.1}$Mn$_{1.5}$In$_{0.5}$     & $I2/m(\alpha0\gamma)00$ & a = 4.471(5) \AA, b = 5.651(7) \AA, c = 4.381(4) \AA\\
                                             &                         & $\beta$ = 92.97(1)$^\circ$, $q$ = 0.339(4)$c^*$\\
\hline
Ni$_{1.8}$Fe$_{0.2}$Mn$_{1.5}$In$_{0.5}$     & $Fm-3m$ ($L2_1$)          & a = 6.012 (2) \AA\\[1ex]
\hline
\end{tabular}
\end{table}

\begin{figure}[h]
\begin{center}
\includegraphics[width=\columnwidth]{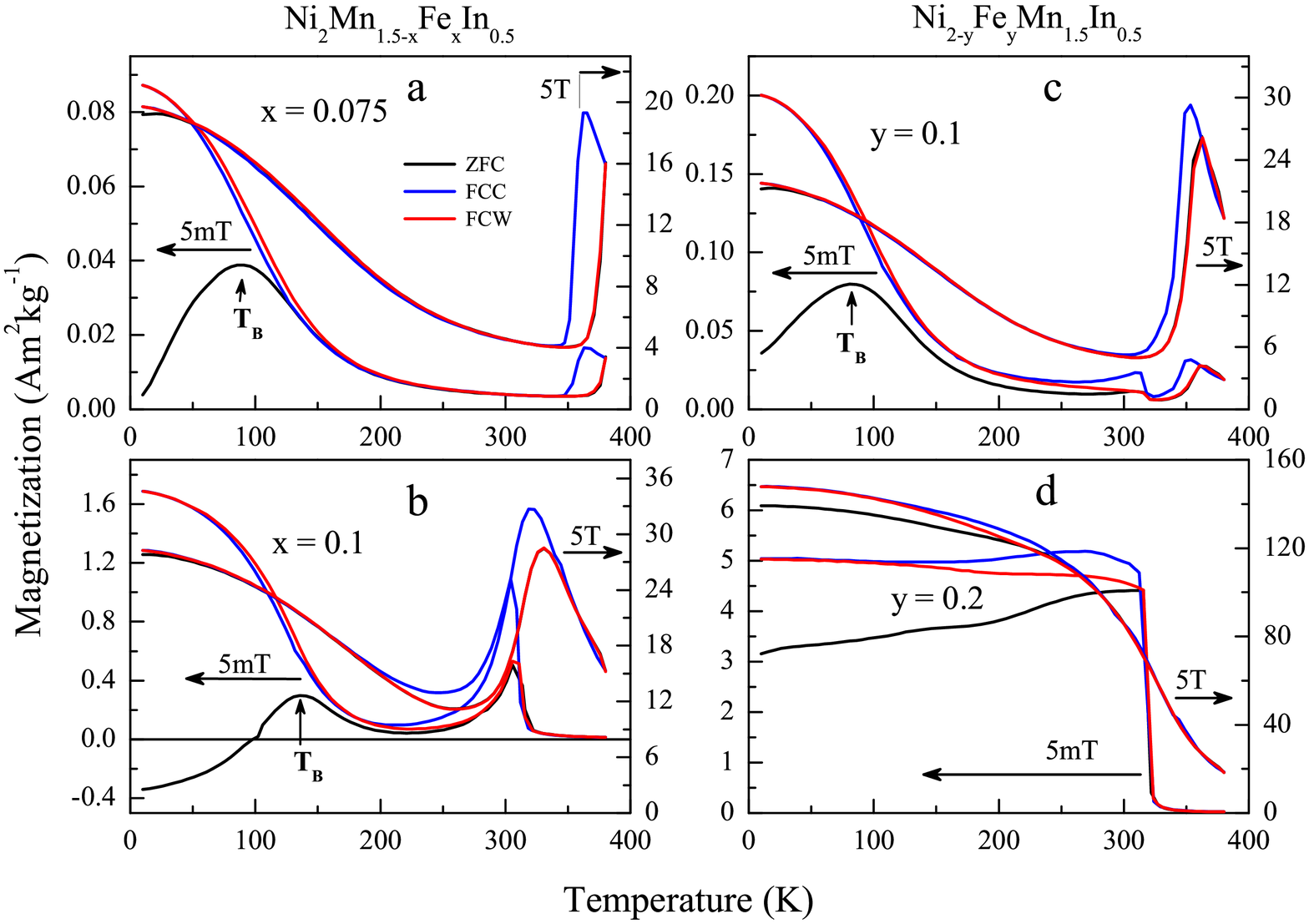}
\caption{Magnetization as a function of temperature for the alloys $x$ = 0.1 and $x$ = 0.2 in the series Ni$_{2}$Mn$_{1.5-x}$Fe$_{x}$In$_{0.5}$ (a, b) and for the alloys $y$ = 0.1 and $y$ = 0.2 in the series Ni$_{2-y}$Fe$_{y}$Mn$_{1.5}$In$_{0.5}$ (c, d) during warming after cooling the alloys in zero field (ZFC) and subsequent warming (FCW) and cooling (FCC) cycles in magnetic field.}
\label{fig:MT}
\end{center}
\end{figure}

The temperature dependent magnetization measurements M(T) recorded in applied magnetic fields of 5 mT and 5T for the alloys $x$ = 0.075 (a), and $x$ = 0.1 (b) belonging to the series Ni$_{2}$Mn$_{1.5-x}$Fe$_{x}$In$_{0.5}$ and for the alloys $y$ = 0.1 (c) and $y$ = 0.2 (d) from the second series Ni$_{2-y}$Fe$_{y}$Mn$_{1.5}$In$_{0.5}$ are shown in Fig. \ref{fig:MT}. The alloys $x$ = 0.075, $x$ = 0.1 and $y$ = 0.1 exhibit a first order martensitic transition evident from the hysteresis in the warming and cooling data. The broad peak at the blocking temperature $T_{B}$  in ZFC data recorded in magnetic field of 5 mT along with the large splitting between ZFC and the field cooled (FC) curves signify a non-ergodic behaviour similar to that observed in Ni$_2$Mn$_{1+x}$In$_{1-x}$ \cite{Nevgi322020}.

\begin{figure}[h]
\begin{center}
\includegraphics[width=\columnwidth]{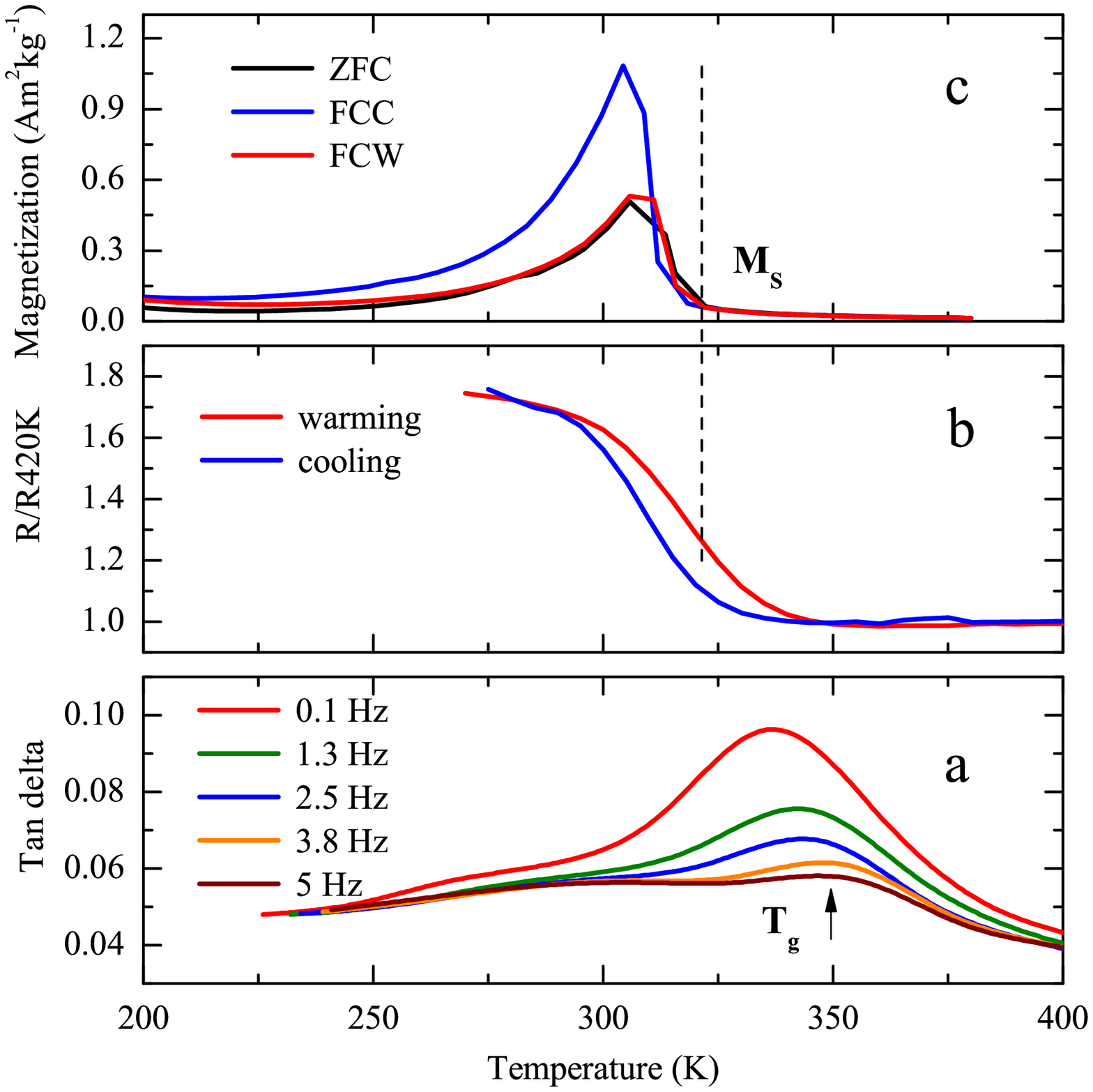}
\caption{Imaginary part of the storage modulus $\tan\delta$ at different frequencies (a), resistivity during warming and cooling cycles (b) and magnetization during ZFC, FCC and FCW cycles recorded in H = 5 mT (c) as a function of temperature for Ni$_2$Mn$_{1.4}$Fe$_{0.1}$In$_{0.5}$}
\label{fig:DMARESMagn}
\end{center}
\end{figure}

In Ni$_{2}$Mn$_{1.5-x}$Fe$_{x}$In$_{0.5}$, as the Fe concentration increases from 0.075 to 0.1, the martensitic transition temperature $T_M$ decreases from 360 K to about 302 K and the $T_{B}$ shifts towards higher temperature, indicating a strengthening of ferromagnetic interactions at the expense of martensitic interactions. Interestingly in $x = 0.1$, the ZFC curve approaches a negative magnetization value below $T$ = 100 K (see Fig. \ref{fig:MT}(b)) signifying a magnetic compensation. It should be noted that the alloy $x = 0.1$ was reported to have a strain glassy ground state below $T_g =$ 350 K \cite{Nevgi1122018}. Frequency dependence of the imaginary part of the storage modulus ($\tan\delta$) seen in Fig. \ref{fig:DMARESMagn}(a) follows Vogel-Fulcher law indicating glassy nature of the ground state. However, as the crystal structure of this alloy ($x = 0.1$) was 7M modulated at room temperature, the strain glass state was classified as unusual one consisting of large undoped martensitic grains separated by an impurity phase. The unconventional nature of the glassy state is also evident from the thermal hysteresis in the resistivity, as shown in Fig. \ref{fig:DMARESMagn}(b). As the hysteresis in magnetization measurement recorded in 5 mT Fig. \ref{fig:DMARESMagn}(c) and resistivity coincide, the observation of the first--order transition in magnetization could be the signature of the martensitic transformation of the large undoped grains. Further, the martensitic transformation of $x = 0.1$ alloy shifts to a higher temperature with an increase in applied field from 5 mT to 5 T (see Fig.\ref{fig:MT}(b)). Such a behaviour of increasing the martensitic transformation  with magnetic field is not seen in the other two transforming alloys, $x = 0.075$ and $y = 0.1$. Contrary to this, the magnetization of Ni$_{1.8}$Fe$_{0.2}$Mn$_{1.5}$In$_{0.5}$ increases sharply displaying a paramagnetic to ferromagnetic transition at $T_C$ = 314 K (Fig. \ref{fig:MT}(d)). However, the presence of hysteresis in the warming and the cooling magnetization curves in the temperature range of 150 K to 350 K need a deeper understanding of the structure.

\begin{figure}[h]
\begin{center}
\includegraphics[width=\columnwidth]{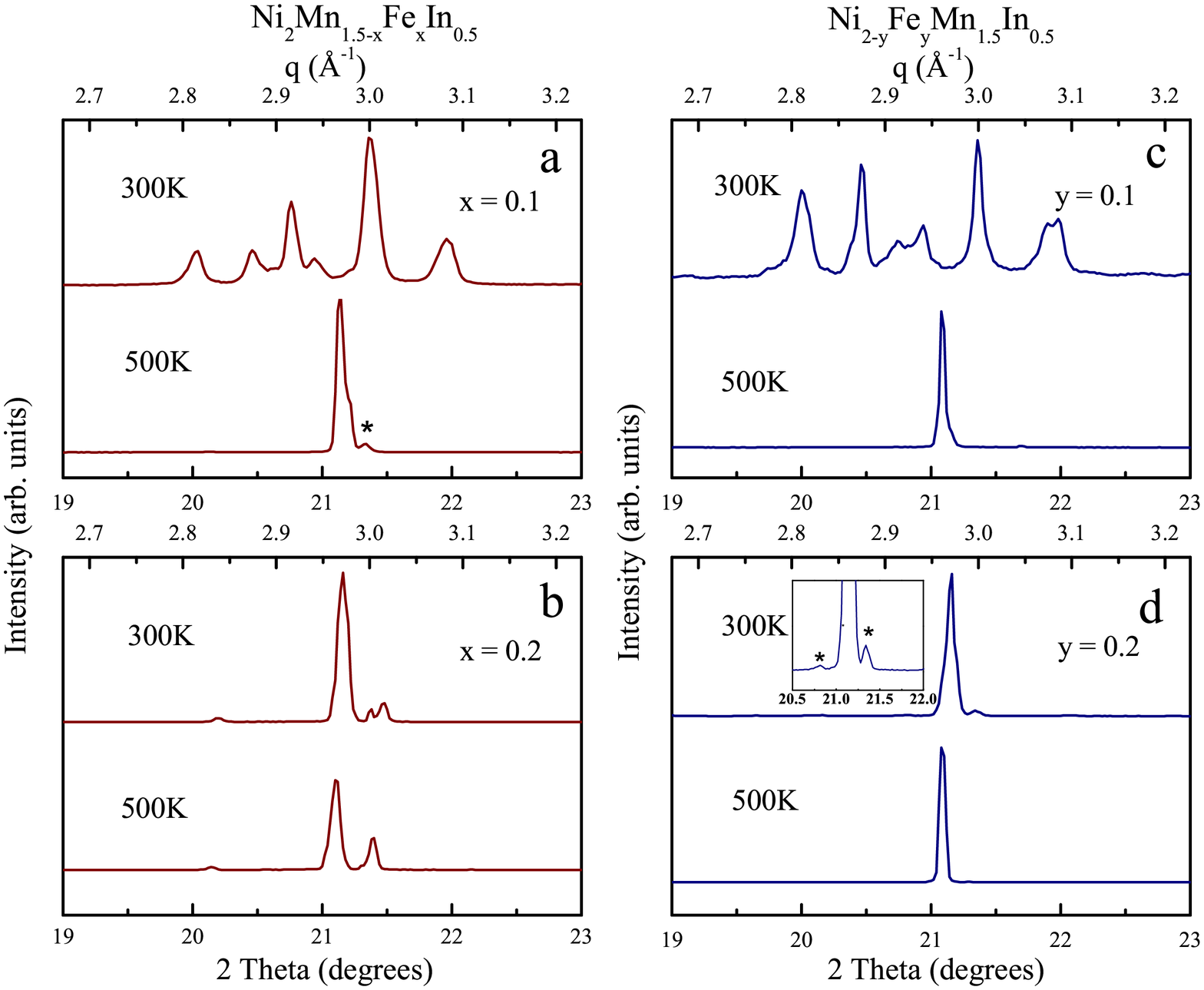}
\caption{X-ray diffraction data in the limited two theta and wave vector q range highlighting the evolution of phases with temperature and Fe concentration in the series Ni$_{2}$Mn$_{1.5-x}$Fe$_{x}$In$_{0.5}$ (a, b) and Ni$_{2-y}$Fe$_{y}$Mn$_{1.5}$In$_{0.5}$ (c, d). The presence of minor impurity phases arising due to Fe substitution are marked as (*).}
\label{fig:SYNCXRD}
\end{center}
\end{figure}

To detect the presence of possible impurity phases as a result of Fe doping in the alloys belonging to Ni$_{2}$Mn$_{1.5-x}$Fe$_{x}$In$_{0.5}$ and Ni$_{2-y}$Fe$_{y}$Mn$_{1.5}$In$_{0.5}$, synchrotron x-ray diffraction measurements were performed at 300 K and at 500 K, which is well above the martensitic transition temperature of the undoped alloy and the phases were identified using the Le Bail refinement. Fig. \ref{fig:SYNCXRD} highlights the diffraction patterns in the limited two theta and wave vector q region. In the series, Ni$_{2}$Mn$_{1.5-x}$Fe$_{x}$In$_{0.5}$,  the alloy $x$ = 0.1 (Fig. \ref{fig:SYNCXRD} a) exhibits 7M modulated martensitic structure at 300 K, which is transformed to a cubic Heusler phase accompanied by a minor fcc impurity phase (marked as $*$) at 500 K. The same impurity phase appears to grow in intensity in $x$ = 0.2 alloy (Fig. \ref{fig:SYNCXRD} b) and is present at both the temperatures along with the major Heusler phase. It appears that the modulated monoclinic structure of the alloy $x$ = 0.1 masked the impurity phase at 300 K, and hence the alloy appeared as a single phase.  The presence of the impurity phase in $x = 0.1$ lends weight to the earlier proposition that the strain glass phase in Ni$_{2}$Mn$_{1.4}$Fe$_{0.1}$In$_{0.5}$ occurs due to large martensitic grains separated by an impurity phase. The martensitic grains transform into austenitic phase above 300K. Since the phase fraction of the impurity phase increases with Fe doping, it either consists of Fe or the impurity phase itself is induced by Fe doping.

In the second series, Ni$_{2-y}$Fe$_{y}$Mn$_{1.5}$In$_{0.5}$ the alloy $y$ = 0.1 (Fig. \ref{fig:SYNCXRD}c) undergoes a transition from the modulated phase at 300 K to a cubic Heusler phase at 500 K. The alloy $y$ = 0.2 (Fig. \ref{fig:SYNCXRD}d) on the other hand exhibits a cubic Heusler phase with a minor percentage of monoclinic phase (I2/m) at 300 K. This monoclinic phase disappears at 500 K, perhaps due to its transformation to the cubic austenitic phase. Same was not seen in the diffraction pattern recorded using a laboratory source, and hence its phase fraction should be meager. The presence of this martensitic impurity phase could be the reason for observed hysteresis in the warming and cooling magnetization curves of $y = 0.2$ alloy. Despite the presence of the impurity phase, these alloys do not exhibit a strain glassy ground state; instead, the martensitic state is completely suppressed by a ferromagnetic cubic phase. Therefore, it appears that the ground state of Fe doped Ni-Mn-In alloys depends on the doping site. In these Mn rich Ni-Mn-In alloys, Ni atoms are expected to occupy the X sites while the Mn atoms are present on the Y sublattice and along with In atoms on the Z sublattices. To determine the site occupancy of the Fe atoms and to study the structural interactions that play a role in the determination of the ground state, we have studied the local environment of Ni, Mn and Fe atoms via EXAFS experiments performed at 300 K and 50 K.

The Ni K and Mn K EXAFS spectra in all alloy compositions have been fitted using  a common structural model based on the crystal structure and similar to that described earlier\cite{Nevgi322020}. It employs 14 independent parameters, including correction to bond length ($\Delta R$) and mean square variation in bond length $\sigma^{2}$ for every scattering path used. The coordination number of the structural correlations were fixed to those obtained for Ni$_{2}$Mn$_{1.5}$In$_{0.5}$ composition. The amplitude reduction factor, $S_0{^2}$ for the two edges were obtained from the analysis of the standard metal spectra and were also kept fixed throughout the analysis.

\begin{figure}[h]
\begin{center}
\includegraphics[width=\columnwidth]{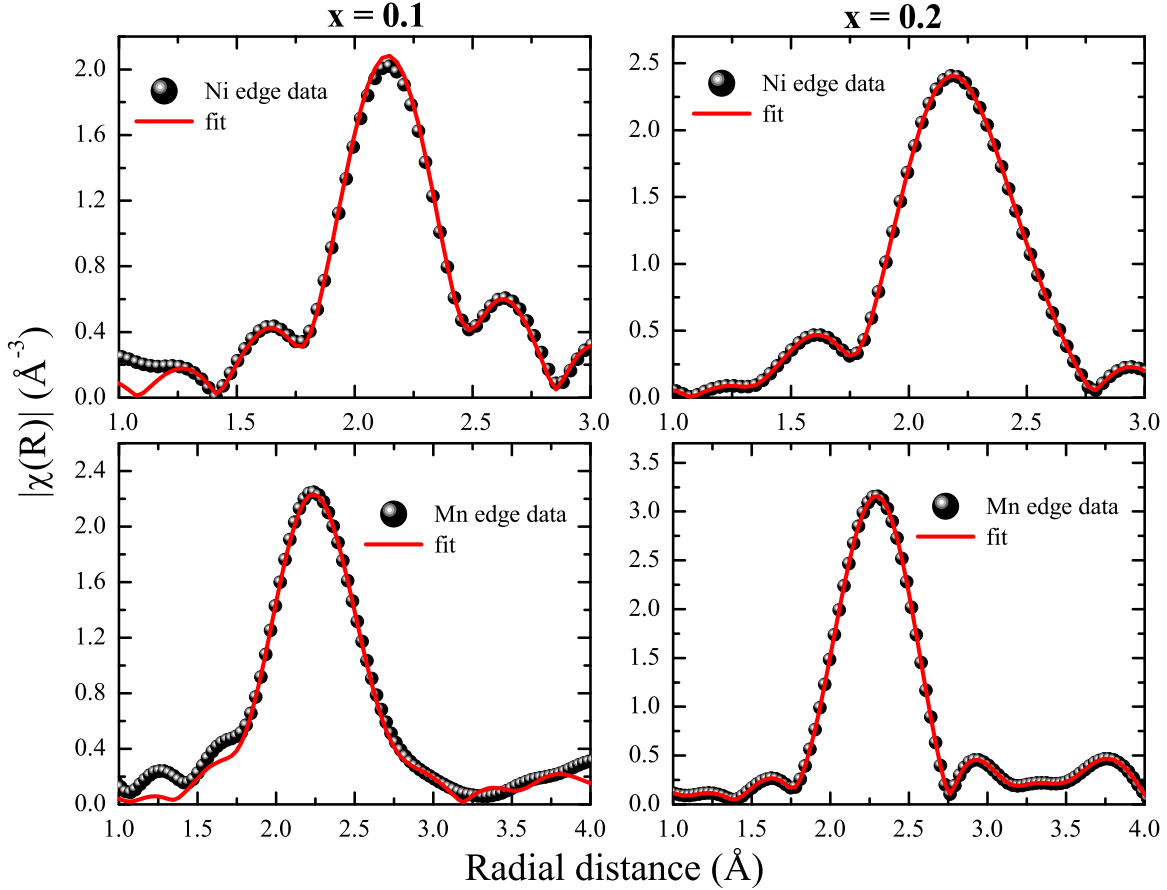}
\caption{Fourier transform magnitude of EXAFS spectra recorded at 50 K at the Ni K and Mn K edges for the alloys $x$ = 0.1 and $x$ = 0.2 in the series Ni$_{2}$Mn$_{1.5-x}$Fe$_{x}$In$_{0.5}$.}
\label{fig:EXAFS1}
\end{center}
\end{figure}

\begin{figure}[h]
\begin{center}
\includegraphics[width=\columnwidth]{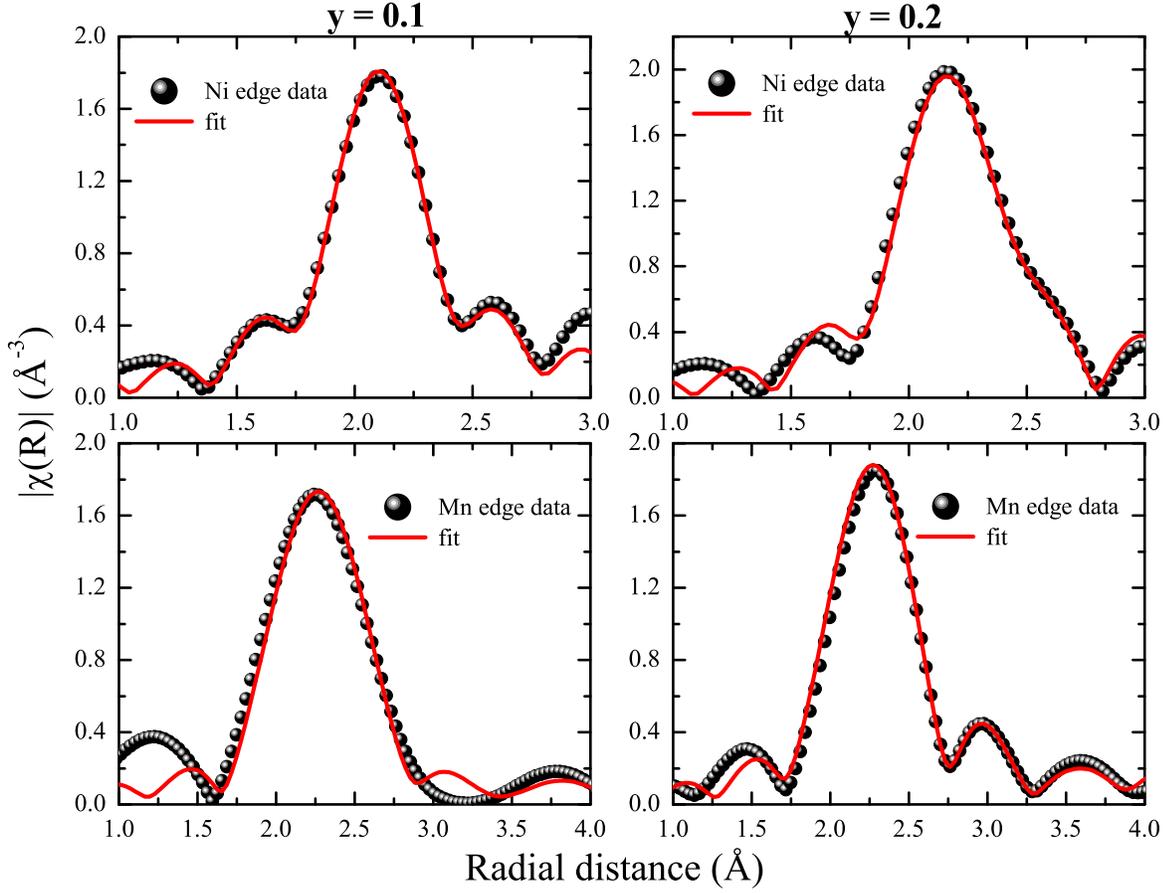}
\caption{Magnitude of Fourier transform of Ni K and Mn K EXAFS spectra for the alloys, Ni$_{2-y}$Fe$_{y}$Mn$_{1.5}$In$_{0.5}$ with $y$ = 0.1 and $y$ = 0.2 at 50 K.}
\label{fig:EXAFS2}
\end{center}
\end{figure}

The best fits for  Ni K and Mn K EXAFS spectra obtained in both the cases, when Fe is doped for Mn and Ni are presented in Fig. \ref{fig:EXAFS1} and Fig. \ref{fig:EXAFS2} respectively and the best fit values of bond length ($R$) and mean square radial displacement ($\sigma^{2}$) at 300 K and 50 K are listed in Table \ref{table1}. The results are in good agreement with previously reported EXAFS studies on such Mn rich Ni-Mn-In alloys \cite{Nevgi322020,Priolkar20113}.  As expected, the nearest neighbour,  Ni--Mn distance is shorter than the  Ni--In bond distance. Similarly, the Mn$_Y$--Mn$_Y$ distance (Mn$_Y$ describes Mn atoms in the Y sublattice of X$_2$YZ Heusler structure) decreases from  $\sim$ 4.4 \AA~ in alloys undergoing a martensitic transformation in the paramagnetic state to $\sim$ 4.2 \AA~ in alloys with dominant ferromagnetic interactions. This is very clearly evident as the Fe content is increased from $y = 0.1$ to $y = 0.2$ in Ni$_{2-y}$Fe$_y$Mn$_{1.5}$In$_{0.5}$, where in the crystal structure at 300 K changes from 7M monoclinic to cubic Heusler along with the strengthening of ferromagnetic interactions. The cubic Heusler structure demands the bond distances Mn--In and Mn$_Y$--Mn$_Z$ (Mn$_Z$ represents Mn atoms occupying Z sublattice) to be equal. This change is also clearly seen as the structure of Fe doped alloys converts from martensitic to cubic Heusler structure.

\begin{figure}[h]
\begin{center}
\includegraphics[width=\columnwidth]{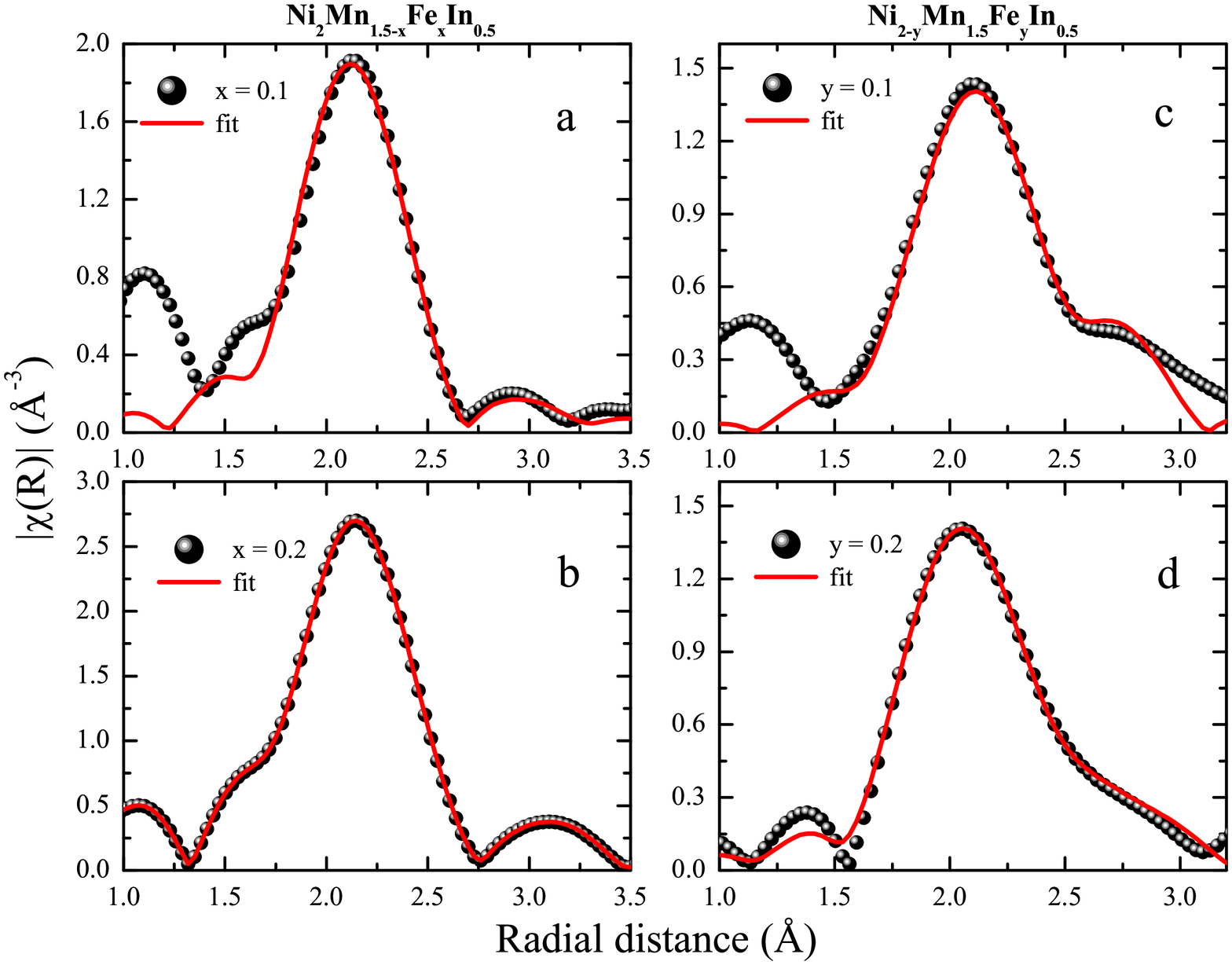}
\caption{Fe K edge EXAFS spectra in $R-$space (Fourier transform magnitude) obtained at 300 K for the alloys Ni$_{2}$Mn$_{1.5-x}$Fe$_{x}$In$_{0.5}$ with (a) $x = 0.1$ and (b) $x= 0.2$ and Ni$_{2-y}$Fe$_{y}$Mn$_{1.5}$In$_{0.5}$ with (c) $y = 0.1$ and (d) $y= 0.2$.}
\label{fig:EXAFS3}
\end{center}
\end{figure}

\begin{table}[h]
\caption{\label{table1} The results of the best fits obtained at 300 K and 50 K at Ni K and Mn K edges in the  $k-$range 3--12 \AA$^{-1}$, and in $R-$range 1--3 \AA. The parameter $R$ gives the bond length while the mean square disorder in the bond length is described by $\sigma^{2}$. Figures in parenthesis indicate uncertainty in the last digit.}
\setlength{\tabcolsep}{1pc}
\vspace{0.3cm}
\begin{adjustbox}{width=\columnwidth,center}
\begin{tabular}{cccccccc}
\hline
\textbf{Alloys}                         &\textbf{Temperature}  &              &\textbf{Bonds}        \\[0.5ex]
                                        &                      &              & Ni - Mn     & Ni - In    &Mn - In    &Mn$_Y$ - Mn$_Y$    &Mn$_Y$ - Mn$_Z$\\
\hline\hline
Ni$_{2}$Mn$_{1.5}$In$_{0.5}$            &{300 K}              & {R (\AA)}            & 2.571(8)	& 2.620(7)	 &2.88(4)    & 4.40(5)   & 3.05(4)  \\
                                        &                     &{$\sigma^{2}$ (\AA$^2$)}       & 0.0057(4)	& 0.0099(8)	 & 0.050(5)  & 0.025(8)  &0.04(3)    \\
\hline
                                        &{50 K}               & {R (\AA)}             &2.58(1)    &2.618(5)    &2.88(4)    &4.45(1)    &3.01(4)   \\
                                        &                     &{$\sigma^{2}$ (\AA$^2$)}        &0.0018(5)  & 0.0068(5)  & 0.008(5)  & 0.02(1)   &0.016(7) \\
\hline
Ni$_{2}$Mn$_{1.4}$Fe$_{0.1}$In$_{0.5}$ &{300 K}              & {R (\AA)}            & 2.57(1)	& 2.62(1)	 &2.86(4)    & 4.39(6)   & 3.05(4)  \\
                                       &                     &{$\sigma^{2}$ (\AA$^2$)}       &0.0105(7)	& 0.005(1)	 & 0.050(5)  & 0.025(8)  &0.04(3)    \\
\hline
                                       &{50 K}               & {R (\AA)}             &2.572(8)   &2.61(1)    &2.89(4)    &4.45(1)    &3.01(4)   \\
                                       &                     &{$\sigma^{2}$ (\AA$^2$)}        &0.001(1)  & 0.0062(5)  & 0.008(5)  & 0.02(1)   &0.016(7) \\
\hline
Ni$_{2}$Mn$_{1.3}$Fe$_{0.2}$In$_{0.5}$ &{300 K}              & {R (\AA)}            & 2.54(1)	& 2.63(1)	 &2.93(3)    & 4.25(7)   & 2.98(4)  \\
                                       &                     &{$\sigma^{2}$ (\AA$^2$)}       & 0.011(1)	& 0.004(1)	 &0.003(4)  & 0.02(1)    &0.010(9)    \\
\hline
                                       &{50 K}               & {R (\AA)}             &2.542(9)    &2.64(2)    &2.94(3)    &4.24(4)    &2.96(2)   \\
                                       &                     &{$\sigma^{2}$ (\AA$^2$)}        &0.002(1)  & 0.0049(6)  & 0.002(2)  & 0.016(4)   &0.003(3) \\
\hline
Ni$_{1.9}$Fe$_{0.1}$Mn$_{1.5}$In$_{0.5}$ &{300 K}              & {R (\AA)}          & 2.577(9)	& 2.633(8)	 &2.88(4)    & 4.39(6)   & 3.05(4)  \\
                                       &                     &{$\sigma^{2}$ (\AA$^2$)}       & 0.0048(8)	& 0.0116(8)	 & 0.050(5)  & 0.025(8)  &0.04(3)    \\
\hline
                                       &{50 K}               & {R (\AA)}             &2.588(7)    &2.619(8)    &2.89(4)    &4.45(1)    &3.02(4)   \\
                                       &                     &{$\sigma^{2}$ (\AA$^2$)}        &0.0015(8)   &0.0072(4)  & 0.008(5)  & 0.02(1)   &0.016(7) \\
\hline
Ni$_{1.8}$Fe$_{0.2}$Mn$_{1.5}$In$_{0.5}$ &{300 K}              & {R (\AA)}           & 2.54(2)	& 2.62(1)	 &2.93(6)    & 4.21(6)   & 2.95(5)  \\
                                       &                     &{$\sigma^{2}$ (\AA$^2$)}       & 0.013(2)	& 0.003(1)	 & 0.026(5)  & 0.02(1)   &0.005(7)    \\
\hline
                                       &{50 K}               & {R (\AA)}             &2.576(7)    &2.625(6)    &2.93(2)    &4.25(5)    &2.95(1)   \\
                                       &                     &{$\sigma^{2}$ (\AA$^2$)}        &0.0080(7)  & 0.0009(5)  & 0.001(1)  & 0.04(4)   &0.08(1)\\ [1ex]
\hline
\end{tabular}
\end{adjustbox}
\end{table}

\begin{table}[h]
\caption{\label{table2} The results of the best fits obtained at 300 K at Fe K edge in the $k-$range 3--12 \AA$^{-1}$, and in $R-$range 1--3 \AA. The parameter $R$ gives the  bond length while the thermal variation in bond length is described by $\sigma^{2}$. Figures in parenthesis indicate uncertainty in the last digit.}
\setlength{\tabcolsep}{1pc}
\vspace{0.3cm}
\centering
\begin{tabular}{ccccc}
\hline
\textbf{Alloys}                           &                  &\textbf{Bonds}        \\[0.5ex]
                                          &                  & Fe - Ni     & Fe - Fe    &Fe - In \\
\hline\hline
Ni$_{2}$Mn$_{1.4}$Fe$_{0.1}$In$_{0.5}$    & {R (\AA)}         & 2.508(8)  	& 3.546(8)	 & \\
                                          &{$\sigma^{2}$ (\AA$^2$)}    & 0.006(3)	    & 0.02(2)	 &  \\
\hline
Ni$_{2}$Mn$_{1.3}$Fe$_{0.2}$In$_{0.5}$    & {R (\AA)}         & 2.525(4)  	& 3.571(4)	 & \\
                                          &{$\sigma^{2}$ (\AA$^2$)}    & 0.009(1)	    & 0.015(6)	 &  \\
\hline
Ni$_{1.9}$Fe$_{0.1}$Mn$_{1.5}$In$_{0.5}$  & {R (\AA)}         & 2.46(5)  	& 2.8(1)	 &2.97(8) \\
                                          &{$\sigma^{2}$ (\AA$^2$)}    & 0.01(1)	    & 0.02(3)	 &0.01(1) \\
\hline
Ni$_{1.8}$Fe$_{0.2}$Mn$_{1.4}$In$_{0.5}$  & {R (\AA)}         & 2.41(2)  	& 2.93(4)	 &3.1(1)\\
                                          &{$\sigma^{2}$ (\AA$^2$)}    & 0.004(3)	    & 0.008(6)	 &0.01(1)\\[1ex]
\hline
\end{tabular}
\end{table}

The Fe K EXAFS spectra recorded at 300 K appears to differ from that of Ni K and Mn K edge spectra in Ni$_{2}$Mn$_{1.5-x}$Fe$_{x}$In$_{0.5}$ indicating a different local environment around Fe  compared to Ni and Mn atoms. Since $x = 0.1$ and $x = 0.2$ compositions reveal presence of an fcc impurity phase, we tried to fit the Fe K edge EXAFS to structural correlations obtained for the fcc structure. A good fit is obtained by considering 12 Ni atoms at $\sim$ 2.5 \AA~ and 6 Fe atoms at $\sim$ 3.5 \AA~ and the same can be seen in Fig. \ref{fig:EXAFS3}(a and b). All attempts to include In atoms as scatterers either in the first shell ($\sim$ 2.5 \AA) or the second shell ($\sim$ 3.5 \AA) did not result in physically acceptable parameters. Therefore, the impurity phase segregated in Ni$_{2}$Mn$_{1.5-x}$Fe$_{x}$In$_{0.5}$ consists of Fe and Ni and from the Fe-Ni binary phase diagram was identified as $\gamma -$(Fe,Ni) phase.

In the second series Ni$_{2-y}$Fe$_{y}$Mn$_{1.5}$In$_{0.5}$, though Fe is doped for Ni, the local structure of Fe is similar to that of Mn at the Y/Z site rather than Ni at the X site. It may be mentioned here that in the Heusler structure, X (Ni) atoms have Y (Mn) and Z (In) atoms in the first coordination shell and X (Ni) atoms in the second coordination shell while the Y (Mn) atoms have only X (Ni) in their first coordination shell and Z (In/Mn) atoms in the second coordination shell. Therefore, if Fe replaces Ni at the X site, then it should have Mn and In atoms as nearest neighbours and Ni atoms as the second nearest neighbours. However, EXAFS signal can be fitted with only Ni atoms in the first coordination shell and In/Mn atoms as second neighbours. The best fit to the experimental data is shown in Fig. \ref{fig:EXAFS3}(c and d), and the parameters obtained from fitting are presented in Table \ref{table2}.

\begin{figure}[h]
\begin{center}
\includegraphics[width=\columnwidth]{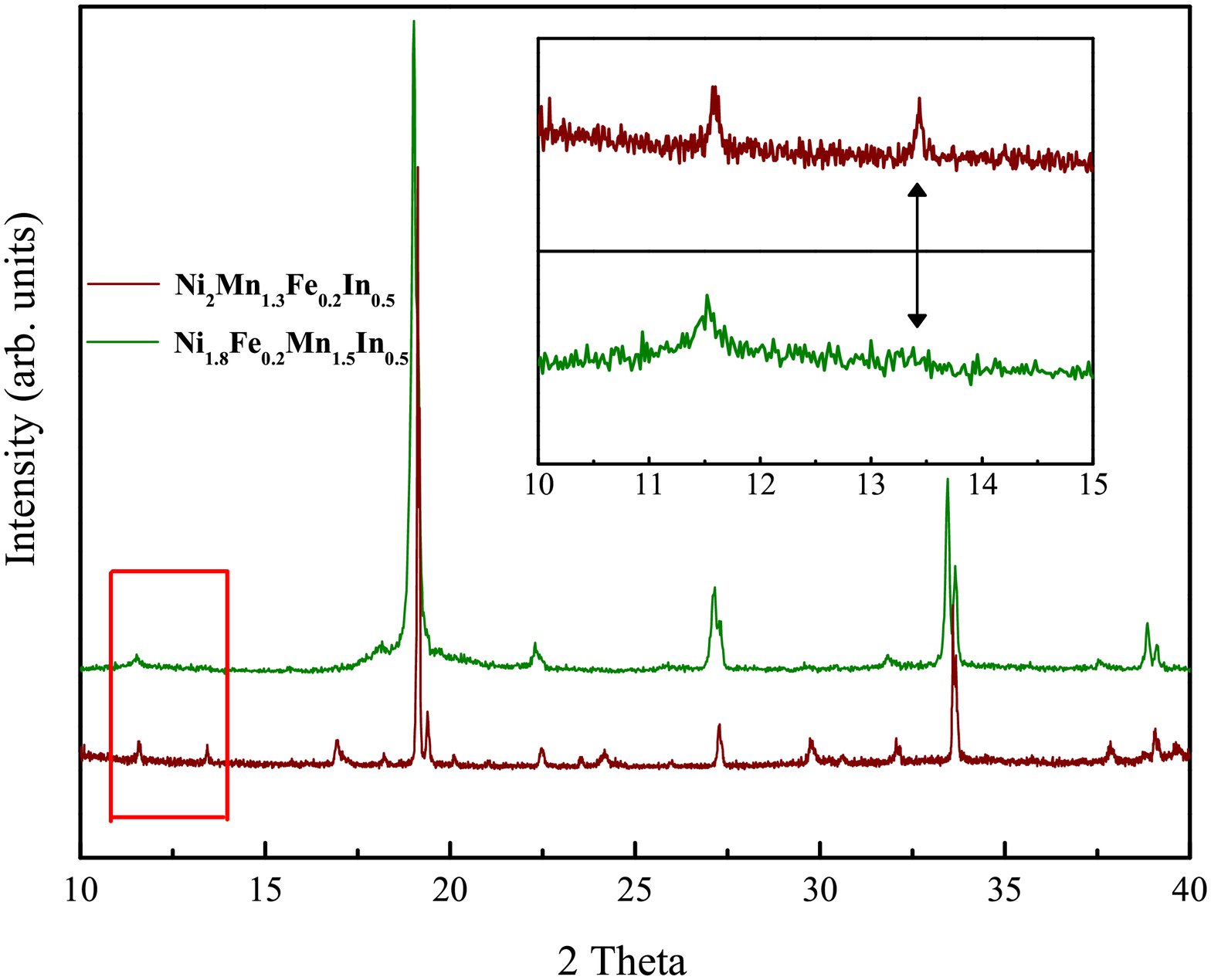}
\caption{A comparison of the x-ray diffraction patterns of the alloys Ni$_{2}$Mn$_{1.3}$Fe$_{0.2}$In$_{0.5}$ and Ni$_{1.8}$Fe$_{0.2}$Mn$_{1.5}$In$_{0.5}$ featuring the absence of 200 peak and the presence of $A2$ disorder in the later.}
\label{fig:XRD1}
\end{center}
\end{figure}

In the X$_2$YZ Heusler structure, the intensities of the super lattice reflections, (111) and (200), are sensitive to the antisite disorder. A disorder in occupancy of Y and Z sublattices of the Heusler structure is identified as $B2$ disorder and results in lowering of intensities of both (111) and (200) reflections. While a disorder involving X and Y sublattices affect the intensity of (200) reflection and is known as $A2$ disorder \cite{Webster101969,Takamura1052009}. Substitution of Fe in the Y/Z sublattice in Ni$_{2-y}$Fe$_{y}$Mn$_{1.5}$In$_{0.5}$ should result in the occupation of the vacant X sites by Mn atoms and therefore result in an $A2$ type disorder. Fig. \ref{fig:XRD1} compares the x-ray diffraction patterns of Ni$_{2}$Mn$_{1.3}$Fe$_{0.2}$In$_{0.5}$ and Ni$_{1.8}$Fe$_{0.2}$Mn$_{1.5}$In$_{0.5}$, highlighting the absence of (200) reflection in Ni$_{1.8}$Fe$_{0.2}$Mn$_{1.5}$In$_{0.5}$ and confirming the presence of $A2$ type disorder.

\section{Discussion}

The above studies illustrate that Fe doping in the martensitic Ni$_{2}$Mn$_{1.5}$In$_{0.5}$ alloy results in the suppression of the martensitic state. The ground state is decided based on whether the dopant atom occupies one of the crystallographic sites in the Heusler structure or segregates in an impurity phase.  These two scenarios are confirmed from both XRD as well as EXAFS studies. In the series Ni$_{2}$Mn$_{1.5-x}$Fe$_{x}$In$_{0.5}$, with the increase in Fe content, an impurity phase, $\gamma -$(Fe,Ni) is segregated in addition to the martensitic Heusler phase. Even though the large Heusler grains undergo martensitic transformation just above room temperature, the ordering of the elastic strain vector is spatially disturbed by the presence of the impurity phase leading to a strain glass phase. This is clear from the fact that the glass transition temperature $T_g = 350$ K is higher than the martensitic transition temperature $T_M =$ 304 K. Furthermore; there is an increase in the $T_M$ of Ni$_{2}$Mn$_{1.4}$Fe$_{0.1}$In$_{0.5}$ alloy from 304 K to 340 K when the applied magnetic field is increased from 5 mT to 5T. In Mn--excess Ni-Mn-In alloys, the martensitic transition temperature generally decreases with magnetic field. This contravening behaviour of increasing $T_M$ under magnetic field can be understood to be due to a strong coupling between magnetic and elastic degrees of freedom. Increased magnetic field lowers the entropy of the system, thereby promoting an increased order of the elastic strain vector and higher transformation temperature.

On the other hand, when Fe is doped for Ni in Ni$_{2-y}$Fe$_{y}$Mn$_{1.5}$In$_{0.5}$, Fe promotes antisite disorder by occupying Y or Z sublattices and forcing Mn to the vacancies in the X sublattice of the Heusler alloy. This $A2$ disorder suppresses martensitic transformation and strengthens ferromagnetic interactions leading to a cubic ferromagnetic ground state. It appears that when the X sublattice in X$_2$YZ Heusler structure is fully occupied, Fe doping results in segregation of an impurity phase, and in the case of Fe being doped into the X sublattice, it promotes antisite disorder and accommodates itself in the Y/Z sublattice of the Heusler structure.

\section{Conclusions}

In conclusion, when Mn is sought to be replaced by Fe in the martensitic Ni$_{2}$Mn$_{1.5}$In$_{0.5}$, the alloy phase separates into a major Heusler phase and a minor, $\gamma -$(Fe,Ni) phase. This $\gamma$ phase serves as an impediment for the long range ordering of the elastic strain vector, promoting a strain glassy ground state. On the other hand, when Fe is added at the expense of Ni, it replaces Mn in the Y/Z Heusler sublattices forcing Mn to occupy the X sublattice along with Ni and thus creating an $A2$  disorder. The presence of antisite disorder suppresses martensitic transition by promoting a ferromagnetic ground state. This study implies that the suppression of martensite via strain glass occurs if the Fe addition facilitates the segregation of an impurity phase. However, if the dopant Fe is accommodated in the Heusler phase even via an antisite disorder; the resultant stronger ferromagnetic interactions forbid martensitic transformation of the resultant alloy.

\section*{Acknowledgements}

KRP and RN acknowledge the Science and Engineering Research Board, Govt. of India under the project SB/S2/CMP-0096/2013 for financial assistance and Department of Science and Technology, Govt. of  India for the travel support within the framework of India\@ DESY collaboration. RN thanks the Council of Scientific and Industrial Research, Govt. of India for Senior Research fellowship. Edmund Welter and Ruidy Nemausat are thanked for experimental assistance at P65 beamline, PETRA III, DESY Hamburg.

\bibliographystyle{iopart-num}
\bibliography{Ref}

\end{document}